\documentclass{PoS}

\usepackage{amsmath,url}
\usepackage{braket}
\usepackage{dsfont}
\usepackage[T1,T5]{fontenc}
\usepackage[utf8]{inputenc}

\newcommand{\MSbar}{\overline{\mbox{MS}}}

\title{Gribov horizon, Polyakov loop and finite temperature}

\ShortTitle{Gribov horizon, Polyakov loop and finite temperature}

\author{Fabrizio Canfora$^{a}$, David Dudal$^{b,c}$, Igor Justo$^{d}$, \speaker{Pablo Pais}$^{e}$\thanks{On leaving from KU Leuven Campus Kortrijk -- Kulak, Department of Physics, Etienne Sabbelaan 53 bus 7657, 8500 Kortrijk, Belgium.}, Luigi Rosa$^{f,g}$, David Vercauteren$^{h}$\\
\llap{$^a$} Centro de Estudios Cient\'{\i}ficos (CECS)\\
Casilla 1469, Valdivia, Chile\\
\llap{$^b$} KU Leuven Campus Kortrijk -- Kulak, Department of Physics\\
Etienne Sabbelaan 53 bus 7657, 8500 Kortrijk, Belgium\\
\llap{$^c$} Ghent University, Department of Physics and Astronomy\\
 Krijgslaan 281-S9, 9000 Gent, Belgium\\
\llap{$^d$} Instituto de Ciencias F\'{\i}sicas y Matem\'{a}ticas Universidad Austral de Chile\\
Casilla 567, Valdivia, Chile \\
\llap{$^e$} Faculty of Mathematics and Physics, Charles University \\
        V Hole\v{s}ovi\v{c}k\'ach 2, 18000 Prague 8, Czech Republic\\
\llap{$^f$} Dipartimento di Matematica e Applicazioni "R. Caccioppoli", Universit\'a di Napoli Federico II \\
Complesso Universitario di Monte S. Angelo Via Cintia Edificio 6, 80126 Napoli, Italia\\
\llap{$^g$} INFN, Sezione di Napoli, Complesso Universitario
di Monte S.~Angelo \\
Via Cintia Edificio 6, 80126 Napoli, Italia\\
\llap{$^h$} Duy T\^an University, Institute of Research and
Development \\
P809, 3 Quang Trung, {\fontencoding{T5}\selectfont H\h ai
Ch\^au, \DJ \`a N\~\abreve ng}, Vietnam\\
Email: \email{fcanforat@gmail.com}, \email{david.dudal@kuleuven.be}, \email{igorfjusto@gmail.com}, \email{pais@ipnp.troja.mff.cuni.cz}, \email{rosa@na.infn.it}, \email{vercauterendavid@dtu.edu.vn}}

\abstract{We consider finite-temperature $SU(2)$ gauge theory in the continuum formulation. Choosing the Landau gauge, the existing gauge copies are taken into account by means of the Gribov-Zwanziger quantization scheme, which entails the introduction of a dynamical mass scale (Gribov mass) directly influencing the Green functions of the theory. Here, we determine simultaneously the Polyakov loop (vacuum expectation value) and Gribov mass in terms of temperature, by minimizing the vacuum energy with respect to the Polyakov-loop parameter and solving the Gribov gap equation. The main result is that the Gribov mass directly feels the deconfinement transition, visible from a cusp occurring at the same temperature where the Polyakov loop becomes nonzero. Finally, problems for the pressure at low temperatures are reported.}

\FullConference{Corfu Summer Institute 2018 "School and Workshops on Elementary Particle Physics and Gravity"\\
		(CORFU2018)\\
		31 August - 28 September, 2018\\
		Corfu, Greece}

\begin{document}

\section{Introduction}
\label{section_Introduction}

Within $SU(N)$ Yang-Mills gauge theories, it is observed that
the asymptotic particle spectrum does not contain the elementary excitations
of quarks and gluons. These color charged objects are confined into color
neutral bound states: this is the so-called color confinement phenomenon. The best theoretical explanation we have for confinement is due to non-perturbative infrared effects. Different criteria for confinement have been proposed (see the pedagogical introduction \cite{Greensite:2011zz}). A very natural observation is that gluons should not belong to the physical spectrum in a
confining theory. Hence, non-perturbative effects should dress the perturbative
propagator in such a way that the positivity conditions are violated. Therefore, it does not belong to the physical spectrum anymore. The Polyakov loop \cite{Polyakov:1978vu} was proposed as an order parameter for the confinement/deconfinement phase transition via its connection to the free energy of a (very heavy) quark. The importance to clarify the interplay between these two
different points of view (nonperturbative Green function's behaviour vs.~Polyakov loop) can be understood by observing that while there
are, in principle, infinitely many different ways to write down a gluon
propagator which violates the positivity conditions, it is very likely that
only few of these ways turns out to be compatible with the Polyakov
criterion.

One of the most fascinating non-perturbative infrared effects is related to the
appearance of Gribov copies \cite{Gribov1978}, that represent an
intrinsic overcounting of the gauge-field configurations which the
perturbative gauge-fixing procedure is unable to take care of.  The
presence of Gribov copies induces the existence of
non-trivial zero modes of the Faddeev-Popov operator, which make the path
integral ill defined. Soon after Gribov's seminal paper, Singer showed that any true gauge condition, as the Landau gauge\footnote{We shall work exclusively with the Landau gauge here.},
presents this obstruction \cite{Singer:1978dk} (see also \cite{Jackiw:1977ng}).

The most effective method to eliminate Gribov copies, at leading order proposed by Gribov
himself, and refined later on by Zwanziger \cite{Gribov1978,Zwanziger1,Zwanziger2,Zwanziger3}, corresponds to restricting the path integral to the {\it first Gribov region},
which is the region in the functional space of gauge potentials over which
the Faddeev-Popov operator is positive definite. The Faddeev--Popov operator is Hermitian in the Landau gauge, so it makes sense to discuss its sign. In \cite{DellAntonio:1989wae,DellAntonio:1991mms} Dell'Antonio
and Zwanziger showed that all the orbits of the theory intersect the Gribov
region, indicating that no physical information is lost when implementing
this restriction. Even though this region still contains copies \cite{vanBaal:1991zw}, this restriction has remarkable
effects. In fact, due to the presence of a dynamical (Gribov) mass scale, the gluon propagator is suppressed while the ghost propagator is enhanced in the infrared. More general, an approach in which the gluon propagator is ``dressed'' by
non-perturbative corrections which push the gluon out of the physical
spectrum leads to propagators and glueball masses in agreement with the
lattice data \cite{Dudal:2010cd,Dudal:2013wja}.

For all these reasons, it makes sense to compute the vacuum expectation value of the Polyakov loop when we eliminate the Gribov copies using the Gribov--Zwanziger (GZ) approach. Related computations are available using different techniques to cope with nonperturbative propagators at finite temperature, see e.g.~\cite{Maas:2011se,Braun:2007bx,Marhauser:2008fz,Reinhardt:2012qe,Reinhardt:2013iia,Heffner:2015zna,Reinosa:2014ooa,Reinosa:2014zta,Fischer:2009gk,Herbst:2013ufa,Bender:1996bm}. In \cite{Zwanziger:2006sc,Lichtenegger:2008mh,Fukushima:2013xsa}, it was already pointed out that the Gribov--Zwanziger quantization offers an interesting way to illuminate some of the typical infrared problems for finite temperature gauge theories.
Therefore, in the present work, we recapitulate the paper \cite{Polyakov_loop}, and present their main results.

\section{The Gribov Ambiguity}
\label{section_Gribov}

In this Section we give a brief overlook of the Gribov ambiguity. The interested reader can find many excellent and fully detailed reviews about this topic (see, for instance, \cite{Sobreiro-Sorella2005,Vandersickel2012}).

Let us start with the Yang--Mills (YM) action in the Euclidean space\footnote{We shall work in the Euclidean space through this work, therefore, we will do not distinguish if the Lorentz indices are up or down.},
\begin{equation}
S_{\text{YM}}\ = \frac{1}{4} \int d^{4}x F^{a}_{\mu\nu}F^{a}_{\mu\nu} \;,  \label{YM_action}
\end{equation}%
which is invariant under local $SU(N)$ gauge transformations
\begin{equation}\label{gauge_transformation}
A_{\mu}\to A'_{\mu}=U^{\dagger}A_{\mu}U+U^{\dagger}\partial_{\mu}U \;,
\end{equation}
where $U\in SU(N)$ is a finite group element.

\begin{figure}
\begin{center}
\includegraphics[width=.6\textwidth]{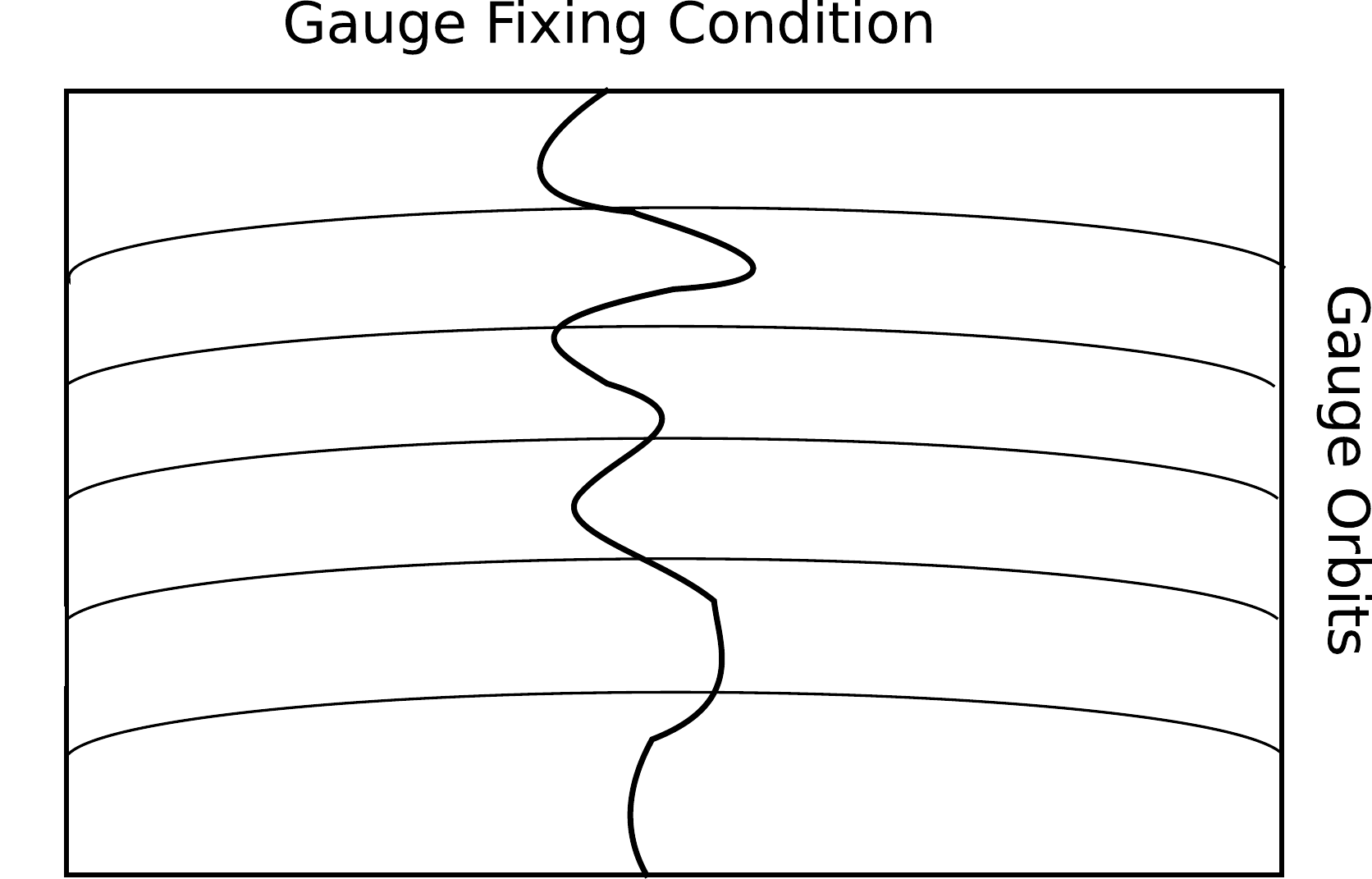}
\end{center}
\caption{An sketched representation of the gauge-fixing condition.}
\label{gauge_fixing_fig}
\end{figure}

When we compute the propagators using path-integral formulation, or any observable of the theory, we must read off the gauge redundance \eqref{gauge_transformation}. In order to do that, we introduce the following unity into the path integral
\begin{equation*}
1=\int \mathcal{D}\alpha \delta(G(A^{\alpha}))\mbox{Det}\left(\frac{\delta G(A^{\alpha})}{\delta \alpha}\right) \;,
\end{equation*}
where $G[A]=\partial_{\mu}A^{\mu}$ is the gauge fixing condition (the Landau gauge in this case), while $A^{\alpha}$ is the infinitesimal gauge transformation \eqref{gauge_transformation} for $g=\mathds{1}-gt^{a}\alpha^{a}$. The proposal of the gauge fixing condition $G[A]$ is to select one (and only one, if possible) representative of the gauge orbit, see Figure \ref{gauge_fixing_fig}.
Thus,
\begin{equation}\label{infinitesimal_gauge_transformation}
A^{\alpha}_{\mu} = - D_{\mu}\alpha^{a} \;,
\end{equation}
where the covariant derivative is $D_{\mu}^{ab}=(\delta ^{ab}\partial _{\mu }-gf^{abc}A_{\mu }^{c})$, $f^{abc}$ being the structure constants of the $SU(N)$ Lie algebra.
Using \eqref{infinitesimal_gauge_transformation}, we can write
\begin{equation*}
\frac{\delta G(A^{\alpha})}{\delta \alpha}=-\partial^{\mu}D_{\mu} \;,
\end{equation*}
Faddeev and Popov decided to represent the determinant of this quantity as
\begin{equation*}
\det\left(-\partial^{\mu}D_{\mu}\right)= \int \mathcal{D}\bar{c}\mathcal{D}c^{b}\exp\left(\bar{c}^{a}\mathcal{M}^{ab}c^{b}\right) \;,
\end{equation*}
where
\begin{equation*}
\mathcal{M}^{ab}=-\partial _{\mu }(\partial _{\mu }\delta
^{ab}-gf^{abc}A_{\mu }^{c}) \;,
\end{equation*}
is the Faddeev--Popov operator. We observe here that $\bar{c},c$ are anticommuting fields, which are scalars under Lorentz transformation, and received the name of {\it Faddeev--Popov ghosts}. Now, the Landau gauge fixing condition can be implemented with the help of an auxiliary field $b$, by adding the following term to the action
\begin{equation*}
-\int d^{4}x b^{a}\partial^{\mu}A^{a}_{\mu} \;.
\end{equation*}

Therefore, in order to avoid over-counting field configurations, we must compute the path integral of the following {\it Faddeev--Popov action}
\begin{equation}\label{FP_action}
S_{\text{FP}} = S_{\text{YM}}-\int d^{4}x\left( b^{a}\partial^{\mu }A_{\mu }^{a}+\bar{c}%
^{a}\partial^{\mu }D_{\mu }^{ab}c^{b}\right)\;.
\end{equation}%

As said before, it is desirable to have only one representative for each gauge orbit. However, this is not possible in the Landau gauge for a non-Abelian Lie algebra, as proven by Gribov \cite{Gribov1978}. Gribov himself proposed a possible solution to this problem by restricting the domain of integration of the Faddeev-Popov (FP) path-integral, showing at the same time that this leads to a gauge and ghost propagators with a modified infrared behaviour. Moreover, these modified gluon propagators could be interpreted as a signal of confinement, as their poles become complex and do not belong to the K\"{a}ll\'{e}n-Lehmann spectrum \cite{Peskin}.

\begin{figure}
\begin{center}
\includegraphics[width=.6\textwidth]{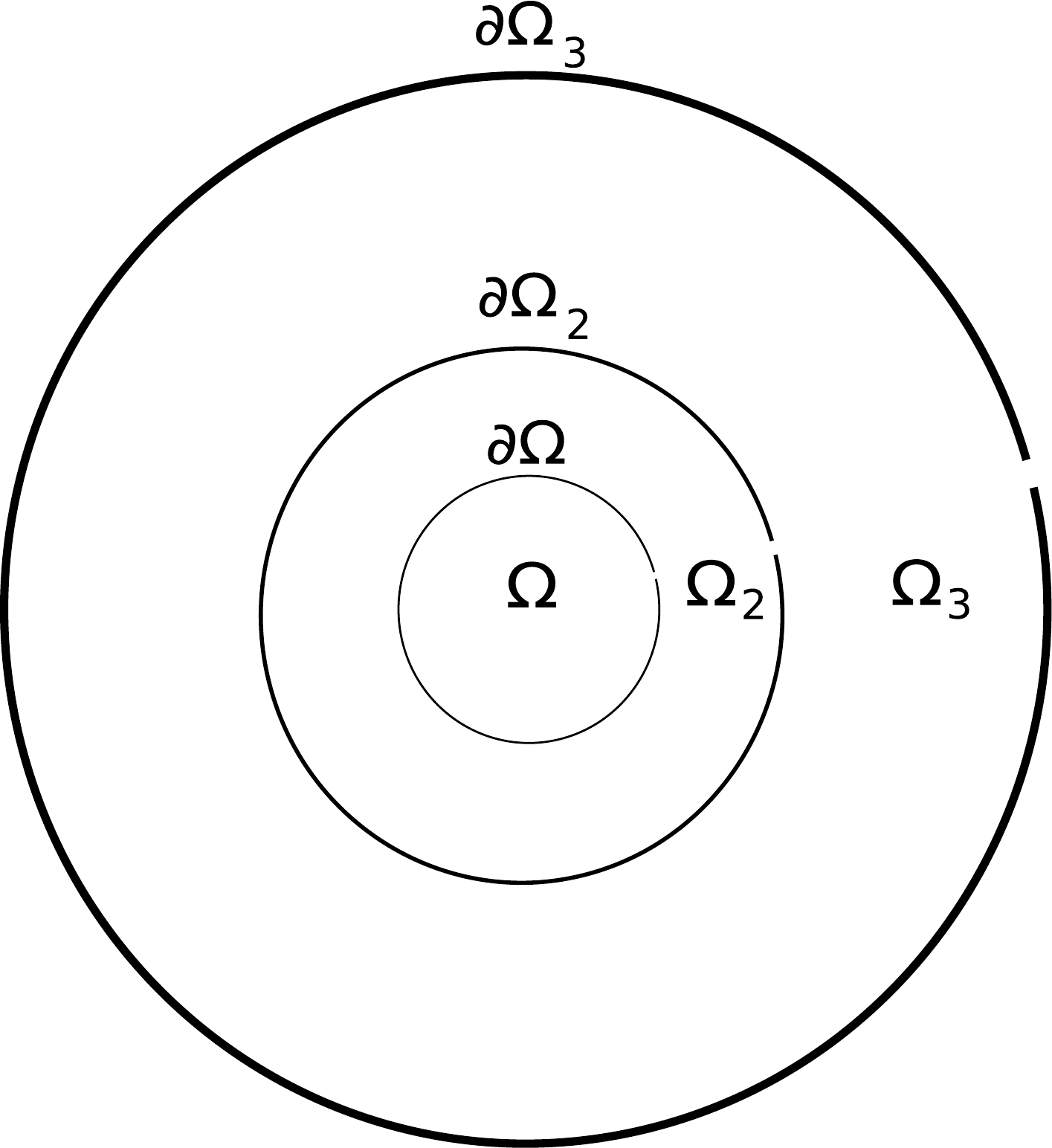}
\caption{The Gribov regions divided by their corresponding Gribov horizons. The first Gribov region is $\Omega$, while in $\Omega_{2}$ $\mathcal{M}^{ab}<0$. In the Gribov horizons $\partial\Omega_{i}$, the FP operator fulfills $\mathcal{M}^{ab}=0$. In $\Omega_{3}$, we have again $\mathcal{M}^{ab}>0$.}
\label{Gribov_region_fig}
\end{center}
\end{figure}

The Gribov region $\Omega$ is the set of all gauge field configurations fulfilling the Landau
gauge, $\partial _{\mu }A_{\mu }^{a}=0$ and $\mathcal{M}^{ab}>0$. Namely,
\begin{equation}\label{Gribov_region}
\Omega \;=\;\{A_{\mu }^{a}\;;\;\;\partial _{\mu }A_{\mu }^{a}=0\;;\;\;%
\mathcal{M}^{ab}>0\;\}\;.
\end{equation}%
As is shown in Figure \ref{Gribov_region_fig}, the space configuration is divided by different horizons $\partial\Omega_{i}$, which are defined by $\mathcal{M}^{ab}=0$.

A great deal of work has been done following the Gribov's seminal paper, remarkably the construction of a renormalizable action, known as Gribov-Zwanziger (GZ) action \cite{Zwanziger1,Zwanziger2,Zwanziger3}. Following these works, the restriction of the domain of integration in the path integral is achieved by adding to the FP action $S_{\text{FP}}$ an additional term $H(A)$, called the {\it horizon term}, given by the following non-local expression
\begin{equation}
H(A,\gamma )={g^{2}}\int d^{4}x\;d^{4}y\;f^{abc}A_{\mu }^{b}(x)\left[
\mathcal{M}^{-1}(\gamma )\right] ^{ad}(x,y)f^{dec}A_{\mu }^{e}(y)\;,
\label{hf1}
\end{equation}%
where $\mathcal{M}^{-1}$ stands for the inverse of the FP
operator. The partition function can then be written as \cite{Gribov1978,Zwanziger1,Zwanziger2,Zwanziger3}:
\begin{equation}
Z_{\text{GZ}}\ =\int_{\Omega }\mathcal{D}A\;\mathcal{D}c\;\mathcal{D}\bar{c}%
\;\mathcal{D}b\;e^{-S_{\text{FP}}}=\int \mathcal{D}A\;\mathcal{D}c\;\mathcal{%
D}\bar{c}\;\mathcal{D}b\;e^{-(S_{\text{FP}}+\gamma ^{4}H(A,\gamma )-V\gamma
^{4}4(N^{2}-1))}\;,  \label{zww1}
\end{equation}%
where $V$ is the Euclidean space-time volume. The parameter $\gamma $ has
the dimension of a mass and is known as the {\it Gribov parameter}. We must stress here the fact that this is not a
free parameter of the theory. It is a dynamical quantity, being determined
in a self-consistent way through a gap equation called the horizon condition
\cite{Gribov1978,Zwanziger1,Zwanziger2,Zwanziger3},
given by
\begin{equation}
\left\langle H(A,\gamma )\right\rangle _{\text{GZ}}\ =4V\left(
N^{2}-1\right) \;,  \label{hc1}
\end{equation}%
where the notation $\left\langle H(A,\gamma )\right\rangle _{\text{GZ}}$
means that the vacuum expectation value of the horizon function $H(A,\gamma )
$ has to be evaluated with the measure defined in Eq.\eqref{zww1}. An
equivalent all-order proof of eq.\eqref{hc1} can be given within the
original Gribov no-pole condition framework \cite{Gribov1978}, by looking
at the exact ghost propagator in an external gauge field \cite{Capri:2012wx}.

Although the horizon term $H(A,\gamma )$, eq.\eqref{hf1}, is non-local, it
can be cast in local form by means of the introduction of a set of auxiliary
fields $(\bar{\omega}_{\mu }^{ab},\omega _{\mu }^{ab},\bar{\varphi}_{\mu
}^{ab},\varphi _{\mu }^{ab})$, where $(\bar{\varphi}_{\mu }^{ab},\varphi
_{\mu }^{ab})$ are a pair of bosonic fields, while $(\bar{\omega}_{\mu
}^{ab},\omega _{\mu }^{ab})$ are anti-commuting. The partition function $Z_{\text{GZ}}$ in eq.\eqref{zww1} can be
rewritten as \cite{Zwanziger1,Zwanziger2,Zwanziger3}
\begin{equation}
Z_{\text{GZ}}\ =\int \mathcal{D}\Phi \;e^{-S_{\text{GZ}}[\Phi ]}\;,
\label{lzww1}
\end{equation}%
where $\Phi $ accounts for the quantizing fields, $A$, $\bar{c}$, $c$, $b$, $%
\bar{\omega}$, $\omega $, $\bar{\varphi}$, and $\varphi $, while $S_{\text{GZ}%
}[\Phi ]$ is the YM action plus gauge fixing and Gribov--Zwanziger (GZ)
terms, in its localized version,
\begin{equation}
S_{\text{GZ}}\ =S_{\text{YM}}\ +S_{\text{gf}}\ +S_{0}+S_{\gamma }\;,
\label{sgz}
\end{equation}%
with
\begin{equation}
S_{0}=\int d^{4}x\left( {\bar{\varphi}}_{\mu }^{ac}(-\partial _{\nu }D_{\nu
}^{ab})\varphi _{\mu }^{bc}-{\bar{\omega}}_{\mu }^{ac}(-\partial _{\nu
}D_{\nu }^{ab})\omega _{\mu }^{bc}+gf^{amb}(\partial _{\nu }{\bar{\omega}}%
_{\mu }^{ac})(D_{\nu }^{mp}c^{p})\varphi _{\mu }^{bc}\right) \;,  \label{s0}
\end{equation}%
and
\begin{equation}
S_{\gamma }=\;\gamma ^{2}\int d^{4}x\left( gf^{abc}A_{\mu }^{a}(\varphi
_{\mu }^{bc}+{\bar{\varphi}}_{\mu }^{bc})\right) -4\gamma ^{4}V(N^{2}-1)\;.
\label{hfl}
\end{equation}%
It can be seen from \eqref{zww1} that the horizon condition \eqref{hc1}
takes the simpler form
\begin{equation}
\frac{\partial \mathcal{E}_{v}}{\partial \gamma ^{2}}=0\;,  \label{ggap}
\end{equation}%
which is called the gap equation. The quantity $\mathcal{E}_{v}(\gamma)$ is the vacuum energy defined by
\begin{equation}
	e^{-V\mathcal{E}_{v}}=Z_\text{GZ}\;  \label{vce} \;.
\end{equation}

The local action $S_{\text{GZ}}$ in eq.\eqref{sgz} is known as the
{\it Gribov--Zwanziger action}. Remarkably, it has been shown to be renormalizable
to all orders \cite{Zwanziger1,Zwanziger2,Zwanziger3,Maggiore:1993wq,Dudal:2007cw,Dudal:2008sp,Dudal:2010fq,Dudal:2011gd}. This important property of the GZ action is a consequence
of an extenstive set of Ward identities constraining the quantum corrections in general and possible divergences in particular.
In fact, introducing the nilpotent BRST transformations
\begin{eqnarray}
sA_{\mu }^{a} &=&-D_{\mu }^{ab}c^{b}\;,  \notag  \label{brst1} \\
sc^{a} &=&\frac{1}{2}gf^{abc}c^{b}c^{c}\;,  \notag \\
s{\bar{c}}^{a} &=&b^{a}\;,\qquad \;\;sb^{a}=0\;,  \notag \\
s{\bar{\omega}}_{\mu }^{ab} &=&{\bar{\varphi}}_{\mu }^{ab}\;,\qquad s{\bar{%
\varphi}}_{\mu }^{ab}=0\;,  \notag \\
s{\varphi }_{\mu }^{ab} &=&{\omega }_{\mu }^{ab}\;,\qquad s{\omega }_{\mu
}^{ab}=0\;,
\end{eqnarray}%
it can immediately be checked that the GZ action exhibits a soft
breaking of the BRST symmetry, as summarized by the equation
\begin{equation}
sS_{\text{GZ}}\ =\gamma ^{2}\Delta \;,  \label{brstbr}
\end{equation}%
where
\begin{equation}
\Delta =\int d^{4}x\left( -gf^{abc}(D_{\mu }^{am}c^{m})(\varphi _{\mu }^{bc}+%
{\bar{\varphi}}_{\mu }^{bc})+gf^{abc}A_{\mu }^{a}\omega _{\mu }^{bc}\right)
\;.  \label{brstb1}
\end{equation}%
Notice that the breaking term $\Delta$ is of dimension two in the fields.
As such, it is a {\it soft breaking} and the ultraviolet divergences can be controlled at the quantum level. The properties of the soft breaking of the BRST symmetry of the GZ theory and its relation with confinement have been object of intensive investigation in recent years, see
\cite{Baulieu:2008fy,Dudal:2009xh,Sorella:2009vt,Sorella:2010it,Capri:2010hb,Dudal:2012sb,Reshetnyak:2013bga,Cucchieri:2014via,Capri:2014bsa,Schaden:2014bea}. Here, it suffices to mention that the broken identity \eqref{brstbr} is
connected with the restriction to the Gribov region $\Omega $. However, a set of BRST invariant composite
operators whose correlation functions exhibit the K{\"{a}}ll{\'{e}}n-Lehmann
spectral representation with positive spectral densities can be consistently
introduced \cite{Baulieu:2009ha}. These correlation functions can be
employed to obtain mass estimates on the spectrum of the glueballs \cite{Dudal:2010cd,Dudal:2013wja}.

\section{The Polyakov Loop}
\label{section_Polyakov}

In this Section we shall investigate the confinement/deconfinement phase transition
of the $SU(2)$ gauge field theory in the presence of two static sources
of (heavy) quarks. The standard way to achieve this goal is by probing the {\it Polyakov loop} (PL) order parameter,
\begin{eqnarray}
\mathcal{P} = \frac{1}{N} \mbox{Tr} \Braket{P e^{ig\int_{0}^{\beta}dt \;\;
A_{0}(t,x)}}\;,
\end{eqnarray}
with $P$ denoting path ordering, needed in the non-Abelian case to ensure the gauge invariance of $\mathcal{P}$. In this analytical description of the phase transition involving the PL, one usually imposes the so-called {\it Polyakov gauge} on the
gauge field, such that, the time-component $A_{0}$ becomes diagonal and independent of (imaginary) time, meaning that the gauge field belongs to the Cartan subalgebra.

In the $SU(2)$ case, if $\frac{1}{2}g\beta%
\Braket{A_{0}} = \frac{\pi}{2}$ then we are in the ``unbroken symmetry phase''
(confined or disordered phase), equivalent to $\Braket{{\cal P}} = 0$;
otherwise, if $\frac{1}{2}g\beta\Braket{A_{0}} < \frac{\pi}{2}$, we are in the
``broken symmetry phase'' (deconfined or ordered phase), equivalent to
$\Braket{{\cal P}} \neq 0$. Since $\mathcal{P}\propto e^{-F T}$ with $T$ the temperature and $F$ the free energy of a heavy quark, it is clear that in the confinement phase, an infinite amount of energy would be required to actually get a free quark. The broken/restored symmetry referred to is the $\mathbb{Z}_N$ center symmetry of a pure gauge theory (no dynamical matter in the fundamental representation).

Besides the trivial simplification of the PL, when imposing the Polyakov gauge it turns out that the quantity $\Braket{A_{0}}$ becomes a good alternative choice for the order parameter instead of $\mathcal{P}$. This extra benefit can be proven by means of Jensen's
inequality for convex functions and is carefully explained in \cite{Marhauser:2008fz}, see also \cite{Braun:2007bx,Reinhardt:2012qe,Reinhardt:2013iia,Heffner:2015zna,Reinosa:2014ooa}.
As mentioned before, with the Polyakov gauge imposed to the background
field $\bar{A}_{\mu}$, the time-component becomes diagonal and
time-independent. In other words, we have $\bar{A}_{\mu}(x) = \bar{A}%
_{0}\delta_{\mu 0}$, with $\bar{A}_{0}$ belonging to the Cartan subalgebra
of the gauge group. For instance, in the Cartan subalgebra of $SU(2)$ only
the $t^{3}$ generator is present, so that $\bar{A}^{a}_{0} = \delta^{a3}\bar{%
A}^{3}_{0}\equiv \delta^{a3}\bar A_0$. As explained in \cite{Reinosa:2014ooa}, at leading order we then simply find, using the properties of the Pauli matrices,
\begin{equation}
\mathcal{P}=\cos\frac{r}{2}\,,
\end{equation}
where we defined
\begin{equation}
  r=g\beta \bar{A}_0\,,
\end{equation}
with $\beta$ the inverse temperature. Just like before, $r=\pi$ corresponds to the confinement phase, while $0\leq r<\pi$ corresponds to deconfinement. With a slight abuse of language, we will refer to the quantity $r$ as the PL hereafter.

In order to probe the phase transition in a quantized non-Abelian gauge field theory, we use the Background Field Gauge (BFG) formalism, detailed in general in e.g.~\cite{Weinberg:1996kr}. Within this framework, the effective gauge field will be defined as the sum of a classical field $\bar{A}_{\mu}$ and a
quantum field $A_{\mu}$: $a_{\mu}(x) = a_{\mu}^{a}(x)t^{a} = \bar{A}_{\mu}+A_{\mu}$, with $t^{a}$  the infinitesimal generators of the
$SU(N)$ symmetry group. The BFG method is a convenient approach, since the tracking of breaking/restoration of the $\mathbb{Z}_{N}$ symmetry becomes
easier by choosing the Polyakov gauge for the background field.

Within this framework, it is convenient to define the gauge condition for the quantum field,
\begin{eqnarray}
\bar{D}_{\mu}A_{\mu} = 0\;,  \label{LDW}
\end{eqnarray}
 known as the Landau--DeWitt (LDW) gauge fixing condition, where $\bar{D}%
^{ab}_{\mu} =\delta^{ab}\partial_{\mu} - gf^{abc}\bar{A}^{c}_{\mu}$ is the
background covariant derivative. After integrating out the (gauge fixing)
auxiliary field $b^{a}$, we end up with the following YM action,
\begin{eqnarray}
S_\text{BFG} = \int d^{d}x\; \left\{ \frac{1}{4}F^{a}_{\mu\nu}F^{a}_{\mu\nu}
- \frac{\left( \bar{D}A \right)^{2}}{2\xi} + \bar{c}^{a}\bar{D}_{\mu}^{ab}
D^{bd}_{\mu}(a)c^{d} \right\} \;.  \label{bfg}
\end{eqnarray}
Notice that, concerning the quantum field $A_{\mu}$, the condition %
\eqref{LDW} is equivalent to the Landau gauge, yet the action still has  background center symmetry. The LDW gauge is actually recovered in the limit
$\xi \to 0$, taken at the very end of each computation.

The LDW is also plagued by Gribov ambiguities, and the
GZ procedure is applicable also in this instance \cite{Polyakov_loop,Canfora:2016ngn}. Actually dealing with the Gribov copies in the LDW gauge is something very delicate, see \cite{Dudal:2017jfw} and more recently \cite{Kroff:2018ncl}. As we are interested in the qualitative result, we shall take the GZ action modified for the BFG
framework (see \cite{Zwanziger:1982na}) as
\begin{multline}
S_\text{GZ+PLoop} = \int d^{d}x\; \left\{ \frac{1}{4}F^{a}_{\mu\nu}F^{a}_{%
\mu\nu} - \frac{\left( \bar{D}A \right)^{2}}{2\xi} + \bar{c}^{a}\bar{D}%
_{\mu}^{ab}D^{bd}_{\mu}(a)c^{d} + \bar{\varphi}_{\mu}^{ac} \bar{D}%
_{\nu}^{ab}D^{bd}_{\nu}(a) \varphi_\mu^{dc} \right. \\
\left. - \bar{\omega}_{\mu}^{ac} \bar{D}_{\nu}^{ab}D^{bd}_{\nu}(a)
\omega_\mu^{dc} - g\gamma ^{2} f^{abc}A_\mu^a \left( \varphi_\mu^{bc} + \bar{%
\varphi}_\mu^{bc} \right) - \gamma^{4}d(N^{2}-1) \right\}\;.  \label{gzpl}
\end{multline}

\section{The finite temperature effective action}
\label{section_finite_temperature}

Considering only the quadratic terms of \eqref{gzpl}, the integration of the
partition function gives us the following vacuum energy at one-loop order, defined according
to \eqref{vce},
\begin{equation}
\mathcal{E}_{v}=-\frac{d(N^{2}-1)}{2Ng^{2}}\lambda ^{4}+\frac{T}{2V}%
(d-1)\mbox{Tr}\ln \frac{D^{4}+\lambda ^{4}}{\Lambda^{4}}-d\frac{T}{2V}%
\mbox{Tr}\ln \frac{-D^{2}}{\Lambda^{2}}\;,  \label{vace}
\end{equation}%
where $V$ is the Euclidean space volume, $\lambda^{4}=2Ng^{2}\gamma^{4}$, $\gamma$ is the Gribov parameter, $\overline{D}$ is the covariant background
derivative in the adjoint representation defined in \eqref{LDW}, and $\Lambda^{2}$ is a scale
parameter in order to regularize the result.  We stress the fact that in this work we are dealing with $N=2$ colors, although we will frequently
continue to explicitly write $N$ dependence for generality of the formulae. Using the usual Matsubara formalism, we have that $D^{2}=(2\pi nT+rsT)^{2}+\vec{q}^{2}$, where $n$ is the Matsubara
mode, $\vec{q}$ is the spacelike momentum component, and $s$ is the isospin,
given by $-1$, $0$, or $+1$ for the $SU(2)$ case\footnote{The $SU(3)$ case was handled in \cite{Reinosa:2014ooa} as well (see also \cite{Serreau:2015saa}).}.

The general trace is of the form
\begin{equation}  \label{1}
I(m^{2},r,s,T)=\frac{1}{\beta V}\mbox{Tr}\ln \frac{-D^2 + m^2}{\Lambda^{2}} = T \sum_s
\sum_{n=-\infty}^{+\infty}\int\frac{ d^{3-\epsilon}q}{(2\pi)^{3-\epsilon}}
\ln \left(\frac{(2\pi nT + rsT)^2+\vec{q}^2+m^2}{\Lambda^{2}}\right)\,,
\end{equation}
which will be computed numerically.

\begin{figure}
\begin{center}
\includegraphics[width=.5\textwidth]{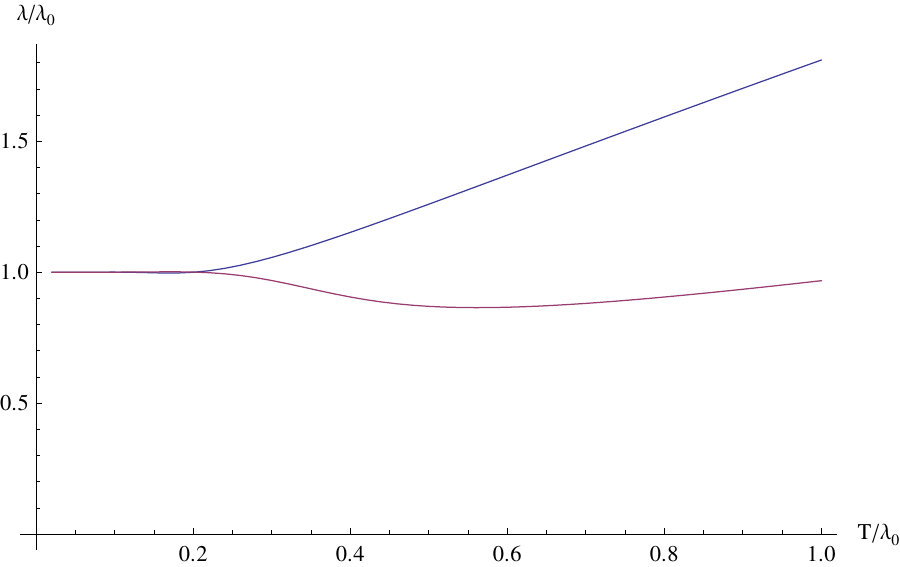}
\end{center}
\caption{The Gribov parameter $\lambda$ as a function of the
temperature $T$ at $r$ equals to zero (upper line) and $\protect\pi$ (lower
line), in units of the zero-temperature Gribov parameter $\protect\lambda_0$. Figure taken from \cite{Polyakov_loop}.}
\label{gammaplot}
\end{figure}

Let us now investigate what happens to the Gribov
parameter $\lambda$ when the temperature is nonzero. Taking the derivative of
the effective potential \eqref{vace} with respect to $\lambda^2$ and
dividing by $d(N^2-1)\lambda^2/Ng^2$ (as we are not interested in the solution $\lambda^2=0$) yields the gap equation for general number of colors $N$:
\begin{equation}  \label{gapeen}
1 = \frac12 \frac{d-1}d Ng^2 \mbox{Tr} \frac1{\partial^4+\lambda^4} + \frac12
\frac{d-1}d \frac{Ng^2}{N^2-1} \frac i{\lambda^2} \sum_s \left(\frac{%
\partial I}{\partial m^2}(i\lambda^2,r,s,T) - \frac{\partial I}{\partial m^2}%
(-i\lambda^2,r,s,T)\right) \;,
\end{equation}
where the notation $\partial I/\partial m^2$ denotes the derivative of $I$
with respect to its first argument. If we
now define $\lambda_0$ to be the solution to the gap equation at $T=0$:
\begin{equation}
1 = \frac12 \frac{d-1}d Ng^2 \mbox{Tr} \frac1{\partial^4+\lambda_0^4} \;,
\label{t0gapeq}
\end{equation}
then we can subtract this equation from the general gap equation %
\eqref{gapeen}. After dividing through $(d-1)Ng^2/2d$ and setting $d=4$ and $%
N=2$, the result is
\begin{equation}
\int \frac{d^4q}{(2\pi)^4} \left(\frac1{q^4+\lambda^4}-\frac1{q^4+%
\lambda_0^4}\right) + \frac i{3\lambda^2} \sum_s \left(\frac{\partial I}{%
\partial m^2}(i\lambda^2,r,s,T) - \frac{\partial I}{\partial m^2}%
(-i\lambda^2,r,s,T)\right) = 0 \;,
\end{equation}
where now all integrations are convergent. This equation can be easily
solved numerically to yield $\lambda$ as a function of temperature $T$ and
background $r$, in units $\lambda_0$. This is shown in \figurename\ \ref{gammaplot}.

\begin{figure}
\begin{center}
\includegraphics[width=.5\textwidth]{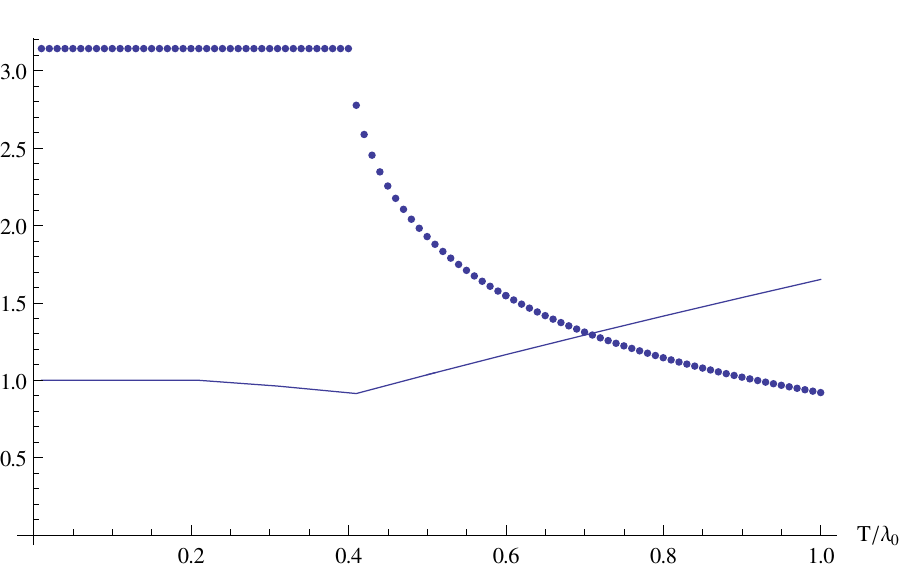}
\end{center}
\caption{The dotted line curve represents $r(T)$, while the continuous line is $\protect\lambda(T)$. At $T\approx 0.40\protect\lambda_{0}$, both curves clearly have a discontinuous derivative. Figure taken from \cite{Polyakov_loop}.}
\label{randlambda}
\end{figure}

Let us now investigate the temperature dependence of $r$. We porpose that the physical value of the background field $r$ is found by minimizing the vacuum energy:
\begin{eqnarray}
\frac{d}{dr} {\cal E}_{v} =0\;.
\end{eqnarray}
From the vacuum energy  \eqref{vace} we have
\begin{eqnarray}
\frac{\partial\mathcal{E}_{v}}{\partial r} = (d-1) \left[ \frac{\partial
I}{\partial r}(i\gamma^2,r,T) + \frac{\partial I}{\partial r}(-i\gamma^2,r,T)
- \frac{d}{(d-1)} \frac{\partial I}{\partial r}(0,r,T) \right] = 0\;.
\label{rgapeq}
\end{eqnarray}
The expression \eqref{rgapeq} was obtained after summation over the
possible values of $s$. Furthermore, we used the fact that $%
I(m^{2},r,+1,T)=I(m^{2},r,-1,T)$ and that $s=0$ accounts for terms
independent of $r$, which are cancelled by the derivation w.r.t. $r$. One can get, whenever $s=\pm1$:
\begin{eqnarray}
\frac{\partial I(m^{2},r,T)}{\partial r} = T \int\frac{d^{3}q}{(2\pi)^{3}}
\frac{2e^{-\frac{\sqrt{\vec{q}^2+m^2}}{T}}\sin r}{\left(1+e^{-2\frac{\sqrt{%
\vec{q}^2+m^2}}{T}}-2e^{-\frac{\sqrt{\vec{q}^2+m^2}}{T}}\cos r \right)}\;.
\end{eqnarray}
Since \eqref{rgapeq} is finite, we can numerically obtain $r$ as a function of
temperature. From the dotted curve in Figure \ref{randlambda} one can easily see that, for $T > T_{\text{crit}} \approx 0.40 \lambda_{0}$, we have $%
r \neq \pi$, pointing to a deconfined phase,  confirming
the computations of the previous section. In the same figure, $\lambda(T)$ is plotted in a continuous line. We observe very clearly
that the Gribov mass $\lambda(T)$ develops a cusp-like behaviour exactly at
the critical temperature $T =T_{\text{crit}}$.

\section{The equation of state issue}

\begin{figure}
\begin{center}
\includegraphics[width=.45\textwidth]{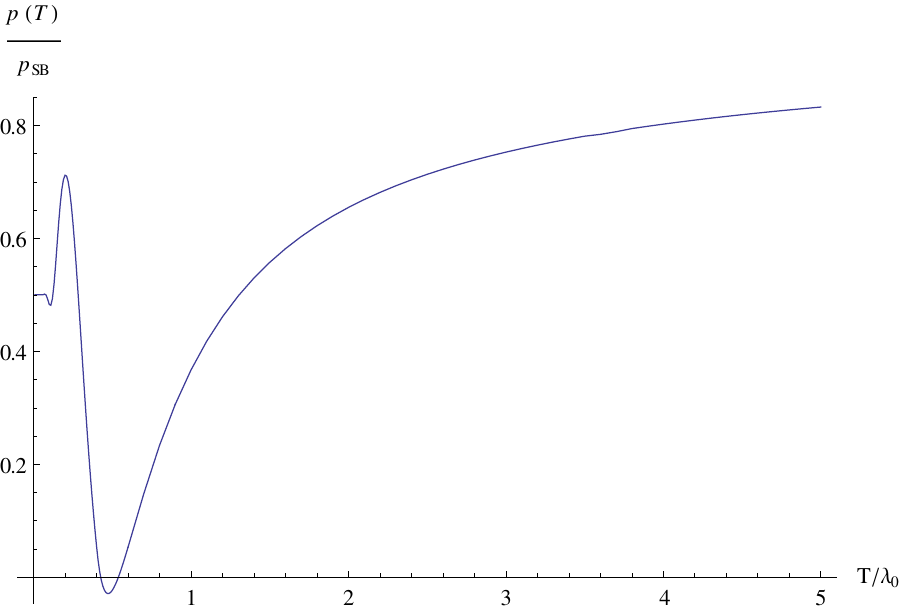} \hspace{10mm} %
\includegraphics[width=.45\textwidth]{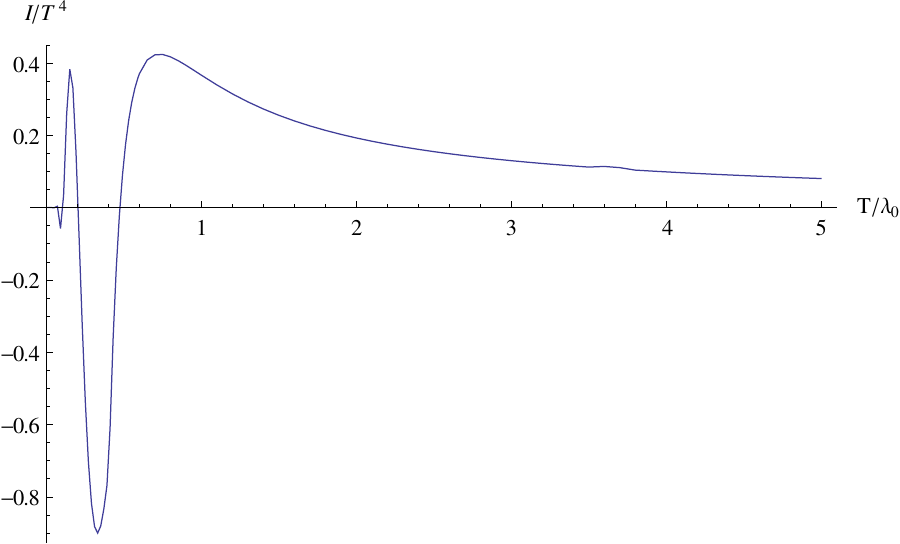}
\end{center}
\caption{Left: GZ pressure (relative to the Stefan-Boltzmann limit pressure $%
\sim T^{4}$). Right: GZ trace anomaly. These plots were taken from \cite{Polyakov_loop}.}
\label{PTraceAnom}
\end{figure}

Following \cite{Philipsen:2012nu}, we can also extract an estimate for the
(density) pressure $p$ and the interaction measure $I/T^{4}$, shown in
Figure \ref{PTraceAnom} (left and right  respectively). As usual the
(density) pressure is defined as
\begin{eqnarray}
p = \frac{1}{\beta V}\ln Z_{GZ}\;,
\end{eqnarray}
which is related to the free energy by $p = -\mathcal{E}_{v}$. Here the plot of the pressure is given relative to the Stefan--Boltzmann limit pressure: $p_{SB} = \kappa T^{4}$, where $\kappa = (N^2 -1)\pi T^{4}/45$ is the Stefan--Boltzmann constant accounting for all degrees of freedom of the system at high temperature. We subtract the zero-temperature value, such that the pressure becomes zero at zero temperature: $p(T) = - [\mathcal{E}_{v}(T) - \mathcal{E}_{v}(T=0)]$. Namely, after using the $\MSbar$ renormalization prescription and choosing the renormalization parameter $\bar{\mu}$ so that the zero temperature gap equation is satisfied,
\begin{eqnarray}
{\bar{\mu}}^2 = \lambda^{2}_{0}e^{-\left(  \frac{5}{6} - \frac{32\pi^2}{3g^{2}} \right)}\;,
\end{eqnarray}
we have the following expression for the pressure (in units of $\lambda_{0}^{4}$),
\begin{eqnarray}
-\frac{p(T)}{\lambda_{0}^{4}} &=& 3 \left[  I(i\lambda'^{2},r,T') + I(-i\lambda'^{2},r,T') - \frac{4}{3}I(0,r,T')\right]
\nonumber \\
&+&
\frac{3}{2} \left[ I(i\lambda'^{2},0,T') + I(-i\lambda'^{2},0,T') - \frac{4}{3}I(0,0,T')\right]
\nonumber \\
&-&
\frac{9\lambda'^{4}}{32\pi^2 } \left(  \ln \lambda'^{2}  - \frac12 \right) - \frac{9}{64\pi^{2}}
\;. \label{vcen1}
\end{eqnarray}
In \eqref{vcen1} prime quantities stand for quantities in units of $\lambda_{0}$, while $\lambda$ and $\lambda_{0}$ satisfy their gap equation. The last term of \eqref{vcen1} accounts for the zero temperature subtraction, so that $p(0) = 0$. Note that the coupling constant does not explicitly appear in \eqref{vcen1} and that $\lambda_{0}$ stands for the Gribov parameter at $T=0$.

The interaction measure $I$ is defined as the trace anomaly in units of $T^{4}$, and $I$ is exactly the trace of the stress-energy
tensor, given by
\begin{eqnarray}
\theta_{\mu\nu} = (p + \epsilon)u_{\mu}u_{\nu} - p\eta_{\mu\nu}\;,
\end{eqnarray}
with $\epsilon$ being the internal energy density, which is defined as $\epsilon = {\cal E}_{v} + Ts$ (with $s$ the entropy density), $u = (1,0,0,0)$ and $\eta_{\mu\nu}$ the (Euclidean) metric of the space-time. Given the thermodynamic definitions of each quantity (energy, pressure and entropy), we obtain
\begin{eqnarray}
I = \theta_{\mu\mu} = T^{5}\frac{\partial}{\partial T}\left(\frac{p}{T^{4}}\right)\;.
\end{eqnarray}
Both quantities display a behavior similar to that presented in \cite{Fukushima:2013xsa} (but note that they plot the temperature in units of the critical temperature ($T_{c}$ in their notation), while we use units $\lambda_{0}$). Besides this, and the fact that we included the effect of PL on the Gribov parameter, in \cite{Fukushima:2013xsa} a lattice-inspired effective coupling was introduced at finite temperature while we used the exact one-loop perturbative expression, which is consistent with the order of all the computations made here.

However, we notice that at temperatures relatively close to our $T_{c}$, the
pressure becomes negative. This is clearly an unphysical feature, possibly
related to some missing essential physics. For higher temperatures, the
situation is fine and the pressure moreover displays a behaviour similar to what is seen
in lattice simulations for the nonperturbative pressure (see \cite{Borsanyi:2012ve} for the $SU(3)$ case). A similar problem is present in one
of the plots presented in \cite[Fig.~4]{Fukushima:2013xsa}, although no comment is made about it. Another strange feature is the oscillating behaviour of both
pressure and interaction measure at low temperatures. Something similar was already observed in \cite{Benic:2012ec} where a quark model was employed with complex conjugate quark mass. It is well-known that the gluon propagator develops two complex conjugate masses in GZ quantization, see e.g.~\cite{Dudal:2010cd,Dudal:2013wja,Baulieu:2009ha,Cucchieri:2011ig} for some more details, so we confirm the findings of \cite{Benic:2012ec} that, at least at leading order, the thermodynamic quantities develop an oscillatory behaviour. We expect this oscillatory behaviour would in principle also be present in \cite{Fukushima:2013xsa} if the pressure and interaction energy were to be computed at lower temperatures than shown there. In any case, the presence of complex masses and their consequences gives us a warning that a certain care is needed when using GZ dynamics, also at the level of spectral properties as done in \cite{Su:2014rma,Florkowski:2015rua}, see also \cite{Baulieu:2008fy,Dudal:2010wn}.

\section{Conclusions}

We have presented here one possible scenario for the confined-deconfined phase transition for a toy model theory, with an internal non-Abelian $SU(2)$ group, where Gribov copies arise. By implementing the PL to the resulting GZ action, gives us a deconfinement critical temperature of $T_{C}\approx0.40\lambda_{0}$.

In the original paper \cite{Polyakov_loop}, it is also investigated how this result is modified if we take into account the condensate of the extra fields, an approach known as {\it refined GZ} (RGZ). The corrected critical temperature in RGZ scenario is $T_{C}=0.250$ $GeV$, which is not far from the lattice value for $SU(2)$ $T_{C}= 0,295$ $GeV$ \cite{Cucchieri:2007ta}. However, we must stress here that we pursuit the quality of the transition behaviour rather than quantify the results, as also the lattice values are getting better tuned as the computation machine capabilities are improved.

The thermodynamic pressure has a negative sector. This is an issue that does not only appear in GZ or RGZ, but it is also shared by many different confinement scenarios, see for instance \cite{Reinosa:2014zta}. Actually, the strange behaviour of the thermodynamic variables is related to the complex poles of the GZ propagator, as is explicitly shown in \cite{CGPRZ_paper}. For that reason, also quark propagators coming from lattice fitting, which are characterized by complex poles, acquire the same pathological thermodynamic behaviour \cite{Benic:2012ec,CGPRZ_paper}. We expect that a more detailed exploration about the trace anomaly both from thermodynamic and hydrodynamic point of view, could shed some light on this issue.

\section*{Acknowledgments}

P.~P. acknowledges the warm hospitality of the Centro de Estudios Cient\'ificos (CECs) in Valdivia, Chile, during different stages of this work. The Centro de Estudios Cient\'ificos (CECs) is funded by the Chilean Government through the Centers of Excellence Base Financing Program of Conicyt.

\bibliographystyle{JHEP}
\bibliography{Corfu2018_biblio}



\end{document}